\documentclass[twoside,english]{elsart}

\usepackage{amssymb}
\usepackage{graphicx}
\usepackage{babel}
\usepackage{multicol}
\usepackage[centerlast]{caption2}
\usepackage{setspace}

\begin{document}

\begin{frontmatter}

\title{Emittance Growth by Beam-Gas Scattering in Single Pass Accelerator}

\author{F. Le Pimpec\corauthref{cor1}},
\ead{Frederic.Le.Pimpec@xfel.eu}
\author{T. Schietinger}
\address{Paul Scherrer Institute \\5232 Villigen
Switzerland}
\corauth[cor1]{Corresponding author}

\begin{abstract}

In order to ensure proper SASE lasing of a 4$^{th}$ Generation light source, one of the goal is to produce
an electron beam of small emittance and to conserve it until its used in the machine undulator.
One of the non recoverable emittance increase is the collision of the electron beam during its transport with the residual gas. Based on previous work by others, we have derived a useful expression of emittance increase for
electrons of any energy and cross check its validity for energies above 100~MeV with an analytical formula.

\end{abstract}


\journal{arxiv.org}

\end{frontmatter}


\section{Introduction}

In a single pass free electron laser accelerator (FEL) one of the
key component is the electron source. The source should provide a
sufficient amount of electrons and should have a low emittance which
has to be preserved to provide the x-ray photons quality requested
by the end users. A small emittance is paramount to reduce the electron energy necessary to lase.
This of course implies a reduced cost for the machine.

Emittance growth happens through multiple channels
\cite{Reiser:1994}. We will only consider emittance growth produced
by the interaction of the electron beam with the residual gas.
\newline During their transport, the electrons will collide with the residual
gas and the emittance will irreversibly increase. The residual gas
pressure in a circular accelerator is usually sufficiently low that this
process is negligible compared to other sources of emittance dilution.
For SwissFEL, a fourth generation light source, the vacuum pressure might not be
of a quality equivalent to that of a storage ring (3rd generation light source),
as the beam will be discarded after use. The quality of the emittance is one of the most important parameters
to obtain lasing \cite{SwissFEL:CDR}.

The pumping scheme of the SwissFEL main Linac employs lump pumps at various location, mainly near the opening of the 2~m C-band and 4~m long S-band RF structures \cite{SwissFEL:CDR}. The dynamic pressure
inside the machine might reach in a worst case scenario the low 10$^{-6}$~Torr range. The question is to understand the quality level of the vacuum required in term of emittance dilution in order to not hamper proper lasing.

\section{Emittance growth after collisions}

Each electron after a collision  with the residual gas will deviate
from its original direction by an angle $\theta$. The following is based on
Raubenheimer's work \cite{Raubenheimer:1991a,Raubenheimer:1991b}.

The single-particle
 invariant before collision is :

\begin{equation}
A^2_x = \gamma x^2 + 2 \alpha x x^{'} + \beta x^{'2}
\label{EquInvartBFR}
\end{equation}

with x(s) the position of the particle, $x^{'}= \frac{dx}{ds}$,
$\beta (s)$ the betatron function, $\alpha(s) = -\frac{1}{2} \frac{d
\beta}{ds}$ and $\gamma (s) = \frac{1 + \alpha(s)^2}{\beta (s)}$.
$A^2_x$ is the emittance.

After a collision equation~\ref{EquInvartBFR} becomes

\begin{equation}
\widetilde{A^2_x} = \gamma x^2 + 2 \alpha x (x^{'} + \Delta x^{'}) + \beta
(x^{'} + \Delta x^{'})^2
\label{EquInvartAFTR}
\end{equation}

From equation~\ref{EquInvartBFR} and equation~\ref{EquInvartAFTR}
and assuming $<$x$> = 0$ and $<$$x^{'}$$> = 0$, we obtain :

\begin{equation}
\Delta A^2_x=  2 \alpha x \Delta x^{'} + 2 \beta x^{'} \Delta x^{'}
+ \beta \Delta x^{'2}
\label{EquInvartAll}
\end{equation}

The change in emittance $\epsilon$ after many collisions :

\begin{eqnarray}
\Delta \epsilon - <\Delta A^2_x> - \beta <\Delta x^{'2}> \ \approx \frac{\beta}{2} <\theta^2_M> \\
\Delta \epsilon \ \approx \frac{\beta}{2} <\theta^2_M>
\label{EquChgEmitt}
\end{eqnarray}

with $<$$\Delta A^2_x$$>$ averaged over the beam and $\theta^2_M$
being the angle after multiple scattering \cite{Reiser:1994}. The
scattering between s and s+ds is then:

\begin{equation}
<\theta^2_M> = N \sigma ds <\theta^2>
\label{EquScattering}
\end{equation}

with N the number of molecules on which the electron scatters,
$\sigma$ the total cross section of collision, which will include
ionization and $\theta$ the single scattering angle. We obtain for
the small angle approximation :

\begin{eqnarray}
<\theta^2> = \frac{1}{\sigma} \int \theta^2 \frac{d \sigma}{d \Omega} (\theta) d\Omega
\ = \frac{2 \pi}{\sigma} \int \theta^3 \frac{d \sigma}{d \Omega} (\theta) d\theta
\label{EquSglScatter}
\end{eqnarray}

and for the emittance, when separating the total cross section and angular
distribution we obtain the following equations.

\begin{eqnarray}
d \epsilon  = \pi N \beta(s) ds \int_0^{\theta_{max}} \frac{d \sigma}{d \Omega} (\theta) \theta^3 d\theta \\
d \epsilon = \pi N \beta(s) \sigma ds \int_0^{\theta_{max}} f1(\theta) \theta^3 d\theta
\label{EquSglScatterEmitt1}
\end{eqnarray}

where $\theta_{max}$ is the maximum scattering angle to be
considered and f$_1$ is the single scattering angular distribution.

\begin{equation}
f_1(\theta) = \frac{1}{\sigma } \frac{d \sigma}{d \Omega} (\theta)
\label{Equf1function}
\end{equation}

The emittance growth is then giving by equation~\ref{EquSglScatterEmitt2}

\begin{equation}
d \epsilon  = \pi N \beta(s) \sigma(E[s]) ds \int_0^{\theta_{max}} f1(\theta,E[s]) \theta^3 d\theta
\label{EquSglScatterEmitt2}
\end{equation}

Where
\begin{itemize}
 \item[*] N: number of gas molecules per cm$^3$.
 \item[*] $\beta (s)$: beta function between s1 and s2.
 \item[*] E(s): electron energy between s1 and s2.
 \item[*] $\sigma (s)$: electron gas cross section in the relevant energy range
 \item[*] f$_1$($\theta$,E): single scattering angular distribution in the relevant
energy range.
 \item[*] Integration of f$_1$ x $\theta^3$ up to $\theta_{max}$, the largest relevant
scattering angle (left to the user to define), as a function of E.
\end{itemize}

\subsection{Gas molecule density}

Using the ideal combined gas law, $N=P/(k_b T)$, at room temperature T=300~K, the number of gas molecules is :

\begin{equation}
N[m^{-3}]  = 3.2 \ 10^{13}  \times P[nTorr]
\label{EquNgasmolec}
\end{equation}

\subsection{Electron cross Section}

In a baked UHV system, the main molecules present are hydrogen and
then carbon based molecules (methane, Carbon mono- and dioxide). The
main culprit for emittance growth is attributed to the elastic
scattering with the gas nuclei over the inelastic scattering
(ionization and excitation) \cite{Raubenheimer:1991b}.

\begin{figure}[htbp]
\centering
\includegraphics[width=0.6\textwidth, clip=]{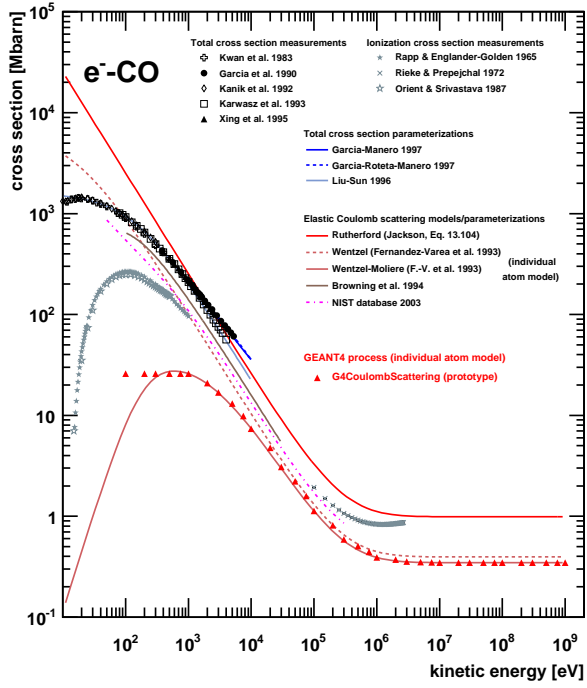}
    \caption{Electron - carbon monoxide (CO) total cross section \cite{Garcia:1997,Garcia2:1997,liusun,Browning:94,Varea:93}. }
\label{figCOcrosssection}
\end{figure}

We will only consider the case of CO, and we will assume that this
is the main gas present in the accelerator beam pipe. With this
assumption, the emittance growth calculated is then overestimated,
as the cross section of collision for CO, see
Fig.\ref{figCOcrosssection} is bigger than for hydrogen.

The total cross section measurement can be found up to 5~keV.
Ionization cross section measurement range from below 1~keV and from
100~keV to 2.7~MeV. The elastic cross section and ionization cross
section have been measured between 10~eV to 10~keV.

The theory Wentzel-Moli\`ere, and other models, expect constant
cross section above 1~MeV \cite{Browning:94,Varea:93}.

The solution adopted for emittance growth estimate is to use the
Browning \cite{Browning:94} parametrization, between 0.1~keV to
30~keV. Below 0.1~keV, we use the Liu-Sun \cite{liusun}
parametrization (for $\sigma_{tot}$)), scaled to match the Browning
curve at 0.1~keV. Above 30~keV, we use the Wentzel-Moliere model,
scaled to match the Browning curve at 30~keV.

For SwissFEL, we will consider the electrons at the exit of the
RF photogun. Their energy is above 6~MeV.
\newline According to Fig.\ref{figCOcrosssection}, the cross section is constant and is
$\sim$1~Mbarn (10$^{-22}$~m$^2$).

\subsection{Betatron function and energy }

\begin{figure}[htbp]
\centering
\includegraphics[width=0.8\textwidth, clip=]{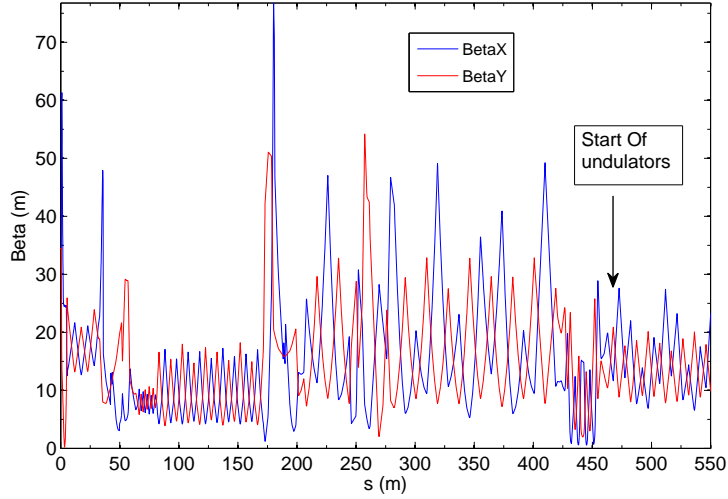}
    \caption{Beta X and Y function along the SwissFEL linac, Aramis beamline }
\label{figBetatronSwiss}
\end{figure}

The betatron function $\beta$(s) and the beam energy E(s) are
calculated along the designed machine using ASTRA and ELEGANT, see Fig.\ref{figBetatronSwiss}
\cite{ASTRA,elegant}. Given the optics we have used for some portion
of the machine an average $\beta$(s) and beam energy E(s).

\subsection{Single Scattering Angular Distribution}

The function f$_1$($\theta$,E)is taken using the " Wentzel-Moli\`ere
" model with the modified screening parameter (due to Moli\`ere),
see Eq.~(26) in Fern\'andez-Varea et al \cite{Varea:93} and
reproduced here.

\begin{eqnarray}
f_1(\theta)  = \frac{1}{\pi} \frac{A(1+A)}{(2A + 1 - cos(\theta))^2} \\
& \mbox{where} \nonumber \\
\lefteqn { A = \left( \frac{\hbar}{2p} \right) ^2 \
\frac{1.13+3.73(\alpha Z/ \beta)^2}{(0.885 \ a_0 \ Z^{-1/3})^2}}
\label{Equf1theta}
\end{eqnarray}

A is the modified screening parameter \cite{Varea:93}, a$_0$ the
classical electron radius, $\alpha$=1/137 the hyperfine constant for
hydrogen, $\beta = v/c$ and Z the atomic number.

\begin{figure}[htbp]
\centering
\includegraphics[width=0.6\textwidth, clip=]{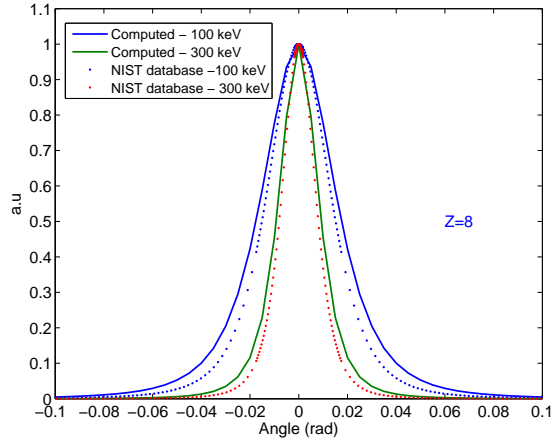}
    \caption{Comparison, for Z=8, between the measured  and calculated distribution of f$_1$($\theta$.)
     Measurement according to equation~\ref{Equf1function}
     and calculated using equation~\ref{Equf1theta}}
\label{figf1NIST}
\end{figure}

Wentzel-Moli\`ere model reproduces well the dependencies of the
angular spread for an electron energy E$>$100~keV when compared to
measurement \cite{NIST:2003}, as shown in Fig.\ref{figf1NIST}.

\begin{figure}[htbp]
\begin{minipage}[t]{.5\linewidth}
\centering
    \includegraphics[width=0.9\textwidth,clip=]{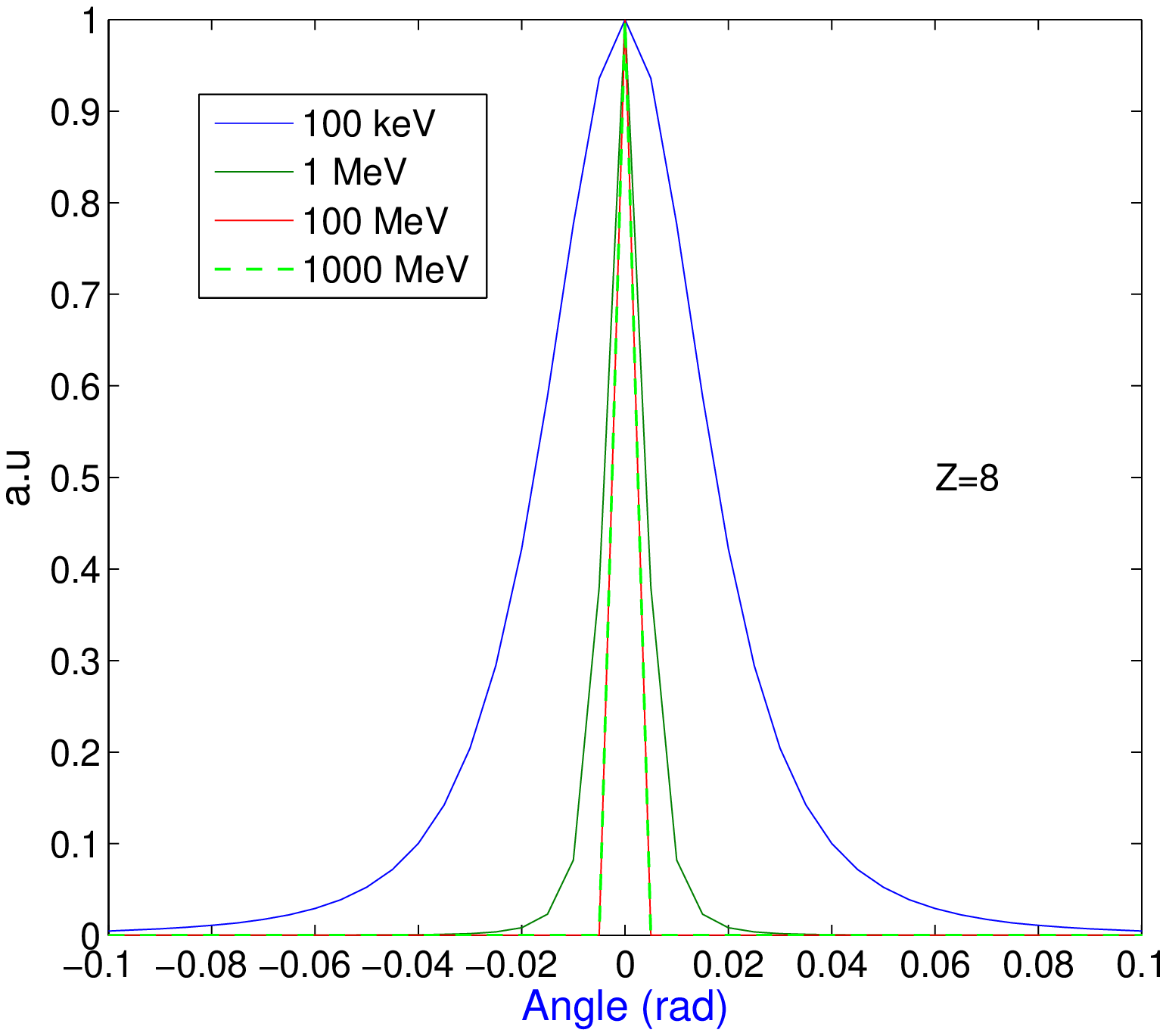}
\end{minipage}%
\begin{minipage}[t]{.5\linewidth}
\centering
    \includegraphics[width=0.9\textwidth,clip=]{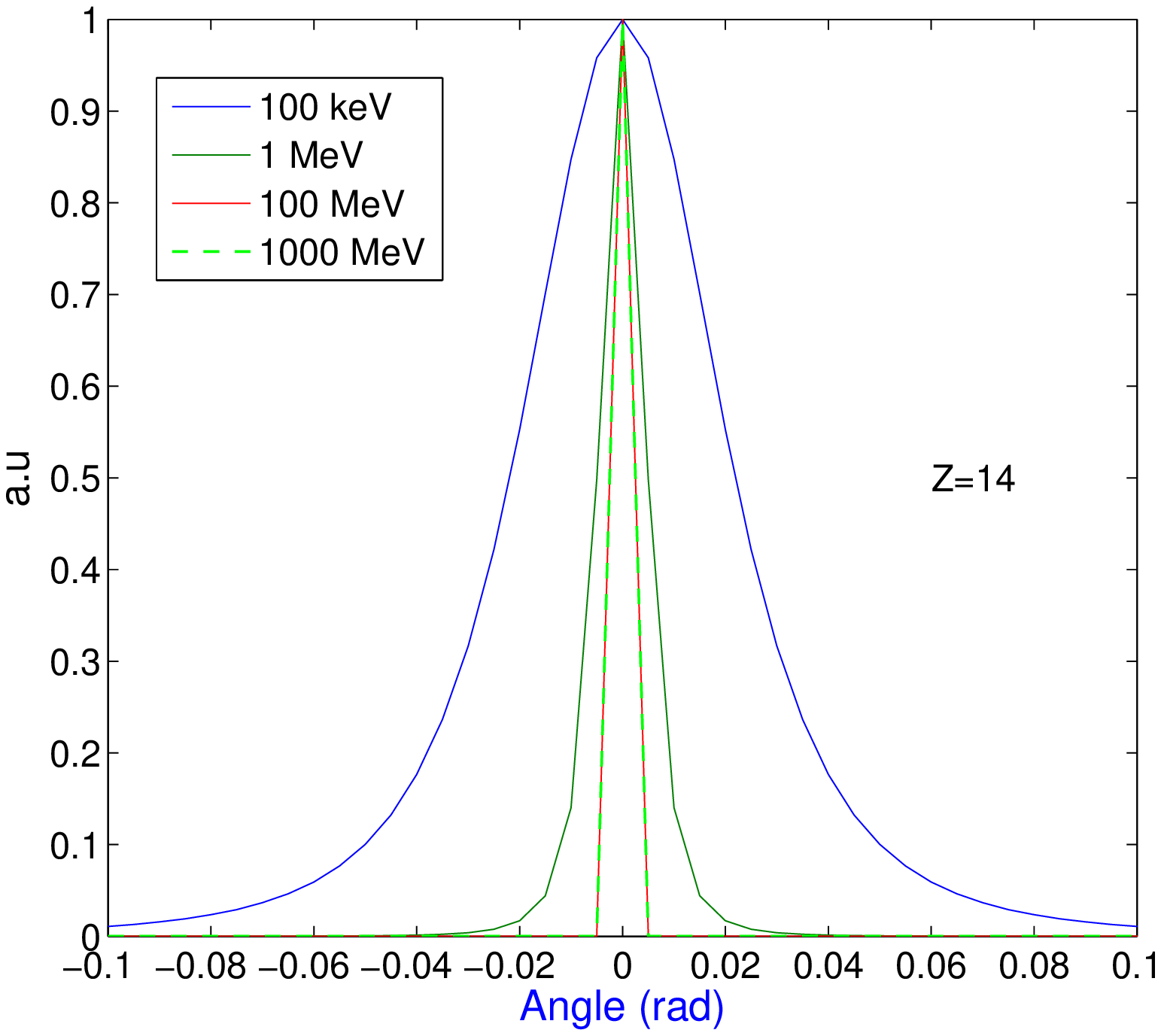}
\end{minipage}
     \caption{Single scattering angular distribution at different electron energies for Z=8 (left figure)
     and Z=14 (right figure) }
\label{figf1theta}
\end{figure}

For E$>$1~MeV, the distribution of f$_1$($\theta$) are very similar
for Z=6, 7, 8 and 14 as shown in Fig.\ref{figf1theta}.

\subsection{Integration of f$_1$.$\theta^3$}

From equation.\ref{EquSglScatterEmitt2}, we now need to integrate
the last term as a function of the energy.

\begin{equation}
I(E)  =  \int_0^{\theta_{max}} f1(\theta,E[s]) \theta^3 d\theta
\label{EquINTf1theta}
\end{equation}

After the second S-band RF structure in SwissFEL, the beam energy E
is above 100~MeV. From Fig.\ref{figf1theta}, we can consider that a
$\theta_{max}$=0.1~rad angle (5.7~deg) is a valid small angle
approximation.

The numerical integration for an energy of 100~MeV and for CO, Z=14 (8 electron for the oxygen and 6 for carbon)
is :

\begin{equation}
I(E=100 \ MeV)  =  1.35 \ 10^{-8}
\label{EquINTf1Numeric}
\end{equation}

For energy above 100~MeV the value for single scattering angular
distribution is smaller as the energy increases, see
equation~\ref{Equf1theta} and shown in Fig.\ref{figf1theta}.

\subsection{Normalized Emittance Increase}

We will overestimate the emittance increase given by
equation.\ref{EquSglScatterEmitt2} by accepting that I(E) is
constant for energies above 100~MeV (equation~\ref{EquINTf1theta}) and is equal to  $1.35 \ 10^{-8}$, equation~\ref{EquINTf1Numeric}.

Equation.\ref{EquSglScatterEmitt2} now becomes :

\begin{equation}
d \epsilon  = \pi \times N [m^{-3}] \times P(nTorr) \times \beta(s) \times \sigma(E[s])
\times I(E(s))
\label{EquEmittINCRease}
\end{equation}

The normalized emittance increase is then given by

\begin{equation}
d \epsilon_n  = (\beta \times \gamma) \times d \epsilon
\label{EquEmittINCReaseNorm}
\end{equation}

where $\beta$ is the ratio of v to the speed of light c, and $\gamma$ is the Lorentz factor
($\gamma$~=~$\frac{1}{\sqrt{(1-\beta^2)}}$).

For electrons having a total energy of 100~MeV ($E~=~\gamma~m_0~c^2$), the lorentz factor $\gamma \sim 200$, $\beta$ can be approximated by taking it equal to 1.

The numeric integration for CO, using a cross section
$\sigma$=0.5~MBarn, Fig.\ref{figCOcrosssection}, a molecular density
N[m$^{-3}$] of 3.2~10$^{13}$, I(E)=9.1~10$^{-9}$
equation.\ref{EquINTf1Numeric}, and an average beta function defined
by the following :

\begin{eqnarray}
<\beta> [m] & = 12.8 \  & \mbox{for} \  0 < s(m) < 170 \nonumber \\
<\beta> [m] & = 19.4 \  & \mbox{for} \  170 < s(m) < 430
\label{Equaveragebeta}
\end{eqnarray}

We obtain for the emittance increase the following  :

\begin{eqnarray}
d \epsilon_{0-170} \  [mm.mrad]  =   3 \ 10^{-7} \  P(nTorr)  \nonumber \\
d \epsilon_{170-430} \  [mm.mrad]   =  6.8 \ 10^{-7} \  P(nTorr)
\label{EquThomasNUM}
\end{eqnarray}

The normalized emittance is
\begin{eqnarray}
d \epsilon_{n(0-170)} \  [mm.mrad]  =  200 \times \ d \epsilon_{0-170} =  6 \ 10^{-5} \  P(nTorr)  \nonumber \\
d \epsilon_{n(170-430)} \  [mm.mrad]   =  200 \times \ d \epsilon_{170-430} = 13.6 \ 10^{-5} \  P(nTorr)
\label{EquThomasNUMNorm}
\end{eqnarray}

Using equation.\ref{EquThomasNUMNorm} for a uniform pressure in the machine of of P$\sim$1000~nTorr, the normalized emittance
increase is :
\begin{eqnarray}
d \epsilon_{n(0-170)} \  [mm.mrad]  =   0.06  \nonumber \\
d \epsilon_{n(170-430)} \  [mm.mrad]   =  0.136
\label{EquEmittCalc}
\end{eqnarray}

The overall emittance increase can be sufficiently significant, given the emittance requirement to obtain lasing. In the case of SwissFEL a normalized emittance better than 0.4~mm.mrad is required.

\section{Emittance growth comparison}

Using the following equation~\ref{EquReiser} for electrons from
Reiser \cite{Reiser:1994}

\begin{equation}
\frac{d \epsilon_n}{ds}  = \frac{8\pi}{k} \ n_s \ \frac{Z_g^2 \
r_c^2}{\beta^3 \ \gamma} \ln(204 \ Z_g^{-1/3})
\label{EquReiser}
\end{equation}

with r$_c$ the classical radius of the electron (2.8 10$^{-15}$~m),
n$_s$ the atoms/m$^3$,  $\beta$ and $\gamma$ the relativistic
factors, Z$_g$ the atomic number of the atom or molecule and k a
wave number. The normalized emittance $\epsilon$ is given in m.rad.

\begin{figure}[htbp]
\centering
    \includegraphics[width=0.66\textwidth,clip=]{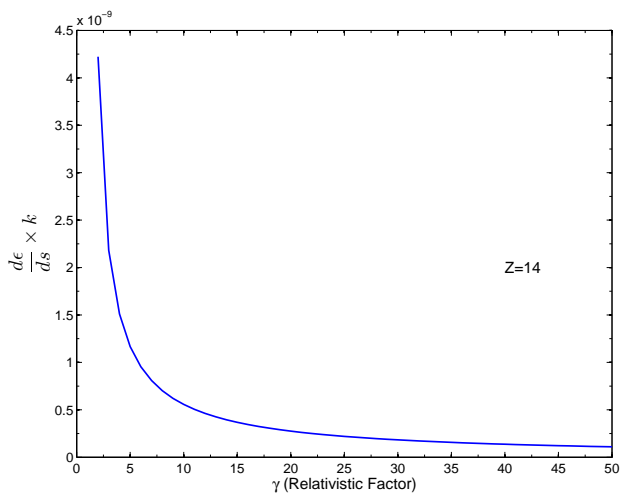}
     \caption{Variation of the normalized emittance, equation.\ref{EquReiser}, in function
     of the particle energy ($\gamma$) for Z=14}
\label{figReiserGamma}
\end{figure}

Before doing the numerical integration it is important to look at the validity of equation.\ref{EquReiser}.
For a total energy of 5~MeV, $\gamma \sim $ 10. Below this value the emittance rises fast and asymptotically to infinity Fig.\ref{figReiserGamma}. This equation models well the emittance increase for highly relativistic electrons. It breaks down at low energy.

The numerical integration for Z$_g$=14 (nitrogen molecule, or carbon monoxide); an electron beam energy of
100~MeV ($\beta \sim 1$ and $\gamma \sim 200$), with a linac length of 170~m, and a pressure of 1~nTorr,
hence n$_s$=3.2 10$^{13}$ at/m$^3$. The average betatron value
$\hat{\beta}$ = 1/k is 12.8~m, see equation~\ref{Equaveragebeta}. We
obtain a normalized emittance (in mm.mrad) increase, normalized to Z=14 :

\begin{equation}
d \epsilon_{n(0-170)} \ [mm.mrad] \sim  5.7 \times 10^{-5} \times (Z/14)^2 \times P[nTorr]
\label{equReiserNUM}
\end{equation}

where Z is the atomic number of the element in the residual gas. For CO, the normalized emittance given by the Reiser \cite{Reiser:1994} equation (equation.\ref{EquReiser}) (0.0607 mm.mrad) is very close to the results obtained using equation.\ref{EquThomasNUMNorm}.

\section{Conclusion}

We have derived a simple equation for an electron beam of any energy. The normalized emittance increased $d \epsilon_n$ can be obtained numerically at each position of the machine. The sum of the emittance increase will give the final increase of the emittance due to collision during the transport of the beam, for the partial pressure of one type of gas. The sum over all types of gases (characterized by their atomic number Z, and their partial pressure in the system) will give the overall emittance increase.

For energies above 100~MeV the analytical equation (Eq.~\ref{EquReiser}) and our solution (Eq.~\ref{EquEmittINCReaseNorm}) are in excellent agreement.
\newline For low beam energy (below 1~MeV), however, care has to be taken in the approximation chosen. First the electron-residual gas cross section varies rapidly, see Fig.\ref{figCOcrosssection}. Second the small angle approximation should be revisited as the scattering angle increases for increasing atomic number ($Z$) and for low beam energy electron, Fig.\ref{figf1theta}. These considerations limit also the application range of equation.\ref{EquEmittINCReaseNorm} due to the approximation taken for its determination. Such considerations would be applicable for DC gun for example \cite{Ganter:2010}.

\section{Acknowledgments}

We would like to thank A. Franchi (ESRF) for his very valuable input and critical look.

%
%


%
%
\clearpage
\listoffigures

\end{document}